\documentclass[3p,times]{elsarticle}

\usepackage{ecrc}
\volume{00}
\firstpage{1}
\journalname{Physics Letters B}
\runauth{A. Ipp et al.}
\jid{plb}
\usepackage{amssymb}
\usepackage[figuresright]{rotating}

\begin{document}

\begin{frontmatter}

\title{Streaking at high energies with electrons and positrons}

\author[label1,label2]{Andreas Ipp}

\author[label1]{J\"org Evers}

\author[label1]{Christoph H.~Keitel}

\author[label1,label3]{Karen Z. Hatsagortsyan}

\address[label1]{Max-Planck-Institut f\"ur Kernphysik, Saupfercheckweg 1, D-69117 Heidelberg, Germany}
\address[label2]{Institut f\"ur Theoretische Physik, Technische Universit\"at Wien, 1040 Vienna, Austria}
\address[label3]{Corresponding author: k.hatsagortsyan@mpi-k.de}

\date{\today}

\begin{abstract}
A detection scheme for characterizing high-energy $\gamma$-ray pulses down to the zeptosecond timescale is proposed. In contrast to existing attosecond metrology techniques, our method is not limited by atomic shell physics and therefore capable of breaking the MeV photon energy and attosecond time-scale barriers. It is inspired by attosecond streak imaging, but builds upon the high-energy process of electron-positron pair production in vacuum through the collision of a test pulse with an intense laser pulse. We discuss necessary conditions to render the scheme feasible in the upcoming Extreme Light Infrastructure laser facility.
\end{abstract}


\end{frontmatter}

\section{Introduction}
\label{intro}
Short photon pulses are an efficient tool for time-resolved monitoring and control of fast-evolving processes. The already well-matured attosecond technique~\citep{Krausz2009} allows  to control the motion of electrons on the atomic scale and to measure inner-shell atomic dynamics with typical energies up to the hundreds of eV and time resolution of several tens of attoseconds~\citep{Paul2001,Hentschel:2001,Sansone,Goulielmakis}. The next challenge of  time-resolving the inner-nuclear dynamics, transient meson states and resonances, or more generally the dynamics of systems governed by the strong interaction~\citep{Ledingham,PHOBOS} requires $\gamma$-rays below attosecond  duration and with energies exceeding the MeV scale \citep{Mourou}. 
A promising example is pump-probe spectroscopy of mesons.
They can be produced from $\gamma$ photons through the Primakoff
effect, i.e., by photoproduction in the Coulomb field of a
nucleus~\citep{PhysRevLett.33.1400,PhysRevC.76.025211}. Various
meson lifetimes fall within the atto- to zeptosecond
regime~\citep{PDG}, e.g., those of $\pi^0$ ($\sim 80$ as), $\eta$
($\sim 0.5$ as), or $\eta'$ ($\sim 3$ zs). Since the most dominant
decay channels of mesons involve photon interactions such as $\eta
\rightarrow \gamma \gamma$ or $\eta \rightarrow \pi^0 \pi^0 \pi^0$
with $\pi^0 \rightarrow \gamma \gamma$~\citep{PDG}, the conversion
of mesons can be enhanced by additional photons. If these photons
arrive within the lifetime of the meson, processes such as $\gamma
\pi^0 \rightarrow \gamma$ can be induced. High-energy photon
double pulses with separation in the attosecond time regime could
thus create (pump) a meson and then enhance or suppress its
possible decay channels and provide information about the
intermediate state of the meson, similar to how pump-probe
experiments explore the evolution of chemical reactions.

Interestingly, there are already suggestions to produce
zeptosecond pulses of keV-energy photons by employing relativistic
laser-plasma interactions~\citep{Pukhov,Bulanov,Nomura}, and short pulses
of multi-MeV energy photons can be produced via nonlinear
Thomson/Compton backscattering~\citep{Hartemann,lan:066501,Kim}. 
At even shorter timescales, there is a proposal for an imploding ultrarelativistic flying mirror which can be created by a megajoule energy laser pulse  at the ultrarelativistic intensity of $10^{24}$ W/cm$^2$ \citep{Tajima,Mourou}. This  would be capable of back-scattering a 10-keV coherent x-ray pulse into a coherent $\gamma$-ray pulse with a duration of 100 ys. Moreover, double pulses of yoctosecond duration of GeV photon energy could be created in non-central heavy ion
collisions~\citep{Ipp:2009ja}.

A basic requirement for the successful application of short $\gamma$-ray pulses is their characterization. Already the accurate measurement of photon pulses emanating from extreme laser field driven plasmas, nuclei, or heavy ion collisions would provide valuable information on the underling physical processes. But at present, no detection schemes are available for the time-dependent characterization of $\gamma$-ray pulses in the MeV--GeV energy range even at moderately short fs-as timescales. 
To achieve attosecond time resolution at lower energy scales, a variety of methods are employed. Autocorrelation schemes use the test pulse and its time-shifted replica (Frequency-Resolved Optical Gating (FROG)~\citep{FROG,Mairesse2005}) or the time- and frequency-shifted replica (Spectral Phase Interferometry for Direct Electric field Reconstruction (SPIDER)~\citep{Iaconis1998,Quere}),  while cross-correlation  schemes are based on the correlation between the test XUV pulse and a femtosecond infrared laser pulse. The latter can be weak, inducing few photon effects (Reconstruction of Attosecond Beating By Interference of
Two-photon Transitions (RABBITT)~\citep{Paul2001}) or strong, yielding attosecond streak imaging \citep{Drescher2001,Itatani2002,Kitzler2002}. Streak imaging~\citep{Drescher2001} is a powerful yet conceptually simple method, in which
a short test pulse (TP) to be characterized is co-propagated with an auxiliary streaking pulse (SP). A nonlinear mechanism  converts photons from the TP to electrons in the presence of the SP. The final momentum distribution of the photoelectrons depends on the phase of the SP at the electron emission moment and hence provides information on the duration and the chirp of the TP. The efficiency of streaking is directly related to the conversion mechanism that depends on the photon energy. For attosecond streak imaging with photon energies of the TP below $100$ eV, the conversion through atomic photoionization is ideally suitable. In the hard x-ray domain, the cross-section of Compton ionization dominates over that of photoionization~\citep{Bethe_Salpeter}, and streak cameras for  hard x-rays can be based on Compton ionization~\citep{Yudin}.
\begin{figure}[b]
\centering
\includegraphics[width=0.4\columnwidth]{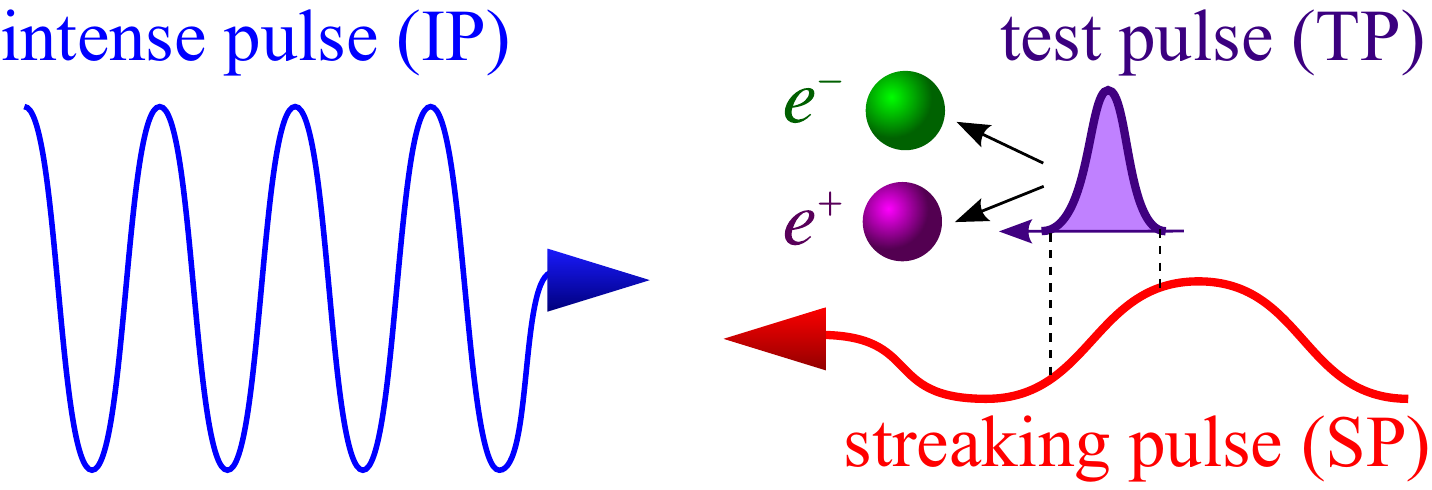}

\caption{\label{fig:1}(color online) Concept of SHEEP. Electron-positron pairs are produced through the interaction of a short test pulse with an intense anti-aligned laser field within a streaking laser pulse. The leptons acquire additional energy and momentum depending on their phase in the streaking pulse at the moment of production.
}
\end{figure}
However, for short pulses of $\gamma$-rays the cross-section of the Compton-effect is significantly suppressed at energies exceeding the MeV range~\citep{Landau}, decreasing its conversion efficiency. This opens the quest for new schemes capable of characterizing pulses in the sub-attosecond and/or super-MeV regime expected at the projected ELI (Extreme Light Infrastructure) or HiPER (High Power laser Energy Research) facilities.

In this Letter, we propose a detection scheme for the
characterization of short $\gamma$-ray pulses of super-MeV energy
photons down to the zeptosecond scale, which we call in the
following ``Streaking at High Energies with Electrons and
Positrons'' (SHEEP). The basic concept of SHEEP is shown in
Fig.~\ref{fig:1}. It is based on the  strong field
electron-positron pair production from vacuum from a
$\gamma$-photon of the TP, assisted by an auxiliary
counter-propagating intense laser pulse (IP). 
In contrast to conventional streak imaging, two particles with opposite charges, electron and positron, are created in the same relative phase within a SP that co-propagates with the TP. By measuring simultaneously the momentum and energy of electrons and positrons originating from different positions within the TP, its length and, in principle, its shape can be reconstructed. 
We analyze SHEEP for different classes of TPs from fs to zs duration, and discuss effects limiting the resolution and the detectable photon energies. 
While pair production by a $\gamma$-photon in a strong laser field at the threshold of the nonperturbative regime has already been observed experimentally in a benchmark experiment at
SLAC~\cite{PhysRevLett.79.1626}, once realized, SHEEP could be the
first viable application for this strong field QED process \cite{PP},
exploiting it as a measurement tool.

\section{The SHEEP concept and the realization conditions}
\label{concept}

The SHEEP concept (see Fig.~\ref{fig:1}) is realized in the collision of three photon pulses with specific functions. The $\gamma$-photons of the TP under characterization, with a  $4$-momentum $k_t$, collide with a counter-propagating infrared IP  of a linear polarization and are converted into electron-positron pairs. The indices of ``$t$'', ``$s$'', or ``$i$'' will be used to refer to the TP, SP, or IP, respectively. The SP  co-propagates with the $\gamma$-ray pulse and is linearly polarized. For simple streaking dynamics, the polarization of the SP and IP are chosen to be perpendicular, with IP polarized along the $x$- and the SP along the $y$-axis. The number of photons in the corresponding pulse is denoted by $N_j$, the pulse duration by $\tau_j$, and the photon energy by  $\omega_j$ ($j\in\{t,s,i\}$, $\hbar=c=1$ units are used throughout).

The first requirement for SHEEP is that a sufficient number of
electron-positron pairs is created by the laser fields. The strong
field pair production process is governed by two relativistic
invariant parameters $\xi=e \sqrt{A_{\mu}A^{\mu}}/m$ and
$\chi=e\sqrt{(F_{\mu\nu}k_t^{\nu})^2}/m^3$~\citep{Ritus}, where
$A_{\mu}$ and $F_{\mu \nu}$ are the vector potential and the field
tensor of the laser fields, respectively, and $e$ and $m$ are the
absolute value of the charge and the mass of the electron.
The number of pairs (averaged by the photon polarization)
produced during the interaction time $\tau_i$ of the TP photons
with IP is given by~\citep{Ritus}
\begin{equation}
N_{e+e-}=\frac{9\alpha m^2 N_t\tau_i
}{64\omega_t}\left(\frac{\chi}{2\pi}\right)^{3/2}\exp[-8/(3\chi)]\,,
\label{eq:W}
\end{equation}
where $\alpha$ is the fine structure constant. In the chosen geometry
$\chi=(k_ik_t)\xi_i/m^2=2\omega_i\omega_t\xi_i/m^2$
and $\xi^2=\xi^2_i+\xi^2_s$. In our case, $\chi$ depends only on
the field of the counter-propagating IP via $\xi_i$.
This is intuitively understandable, since the role of the
intense laser field in the pair production process by a $\gamma$-photon
is the compensation of the momentum of the $\gamma$-photon which the
co-propagating SP photons cannot fulfill. Therefore,
the characteristic parameter $\chi$ of the process cannot depend on the SP.
The infrared IP should be strong enough to initiate pair production.
Exponential suppression of the pair production probability is avoided if
\begin{equation}
 \chi=\frac{2\omega_i\omega_t\xi_i}{m^2}\gtrsim 8/3\,,
\label{eq:C1}
\end{equation}
which assures that the IP is intense enough to provide the
necessary number of laser photons for the pair production process.
The condition of Eq.~(\ref{eq:C1}) is usually fulfilled when
$\xi_i\gg 1$ which corresponds to the quasi-static limit in which
the probability of strong field QED processes in laser fields
coincides with the one in the crossed field with the same value of
$\chi$~\citep{Ritus}. For $\chi\gtrsim 8/3$, the pair production
process becomes very efficient, and the number of produced pairs
via Eq.~(\ref{eq:W}) is 
\begin{equation}
\label{Nee} N_{e^+e^-}\approx 0.014\alpha
(m^2/\omega_t)N_t\tau_i\,.
\end{equation}
To demonstrate the advantage of SHEEP for high-energy photons
over the alternative scheme of streaking via Compton photoionization~\citep{Yudin}, we
compare the number of  produced pairs $N_{e^+e^-}$
with the number of scattered electrons in the Compton photoionization process
$N_C$.
Using the Compton effect cross-section $\sigma \sim \pi
r_0^2(m/\omega_t)(\ln(2\omega_t/m)+1/2)$~\citep{Landau}, where $Z$
is the atomic number and $r_0=e^2/m$ the classical radius of the
electron, we find $N_C\sim \pi r_0^2 (m/\omega_t)\rho_eN_t\tau_i$,
where  $\rho_e$ is the density of atomic electrons in the Compton
process. Then, $N_C/N_{e^+e^-}\sim 2\times 10^2\alpha\rho_e
\lambdabar_C^3\sim 10^{-8}$ at $\rho_e=10^{23}$ cm$^{-3}$, with
the Compton wavelength
$\lambdabar_C=1/m$. We thus conclude that
the efficiency of the pair production process becomes overwhelming
at high photon energies of TP when $\chi\gtrsim 8/3$ is reached.

The second requirement is that the pair production should be
initiated only by $\gamma$-photons of the TP but not by the SP and the IP.
Therefore, the fields of the SP and IP in the center-of-mass
frame of the electron-positron pairs, hypothetically produced via
the SP and IP, should be negligible with respect to the Schwinger
critical field $E_{cr}=m^{2}/e$ ~\citep{Ritus}. The center-of-mass
frame is determined by the equality of the Doppler-shifted
frequencies of the SP and IP,
$2\gamma_{cm}\omega_i=\omega_s/2\gamma_{cm}$, with the
Lorentz-factor of the center-of-mass frame $\gamma_{cm}$. The
conditions for the suppression of the pair production by the SP
and IP interaction, $2\gamma_{cm}E_i\ll E_{cr}$ and
$E_s/2\gamma_{cm}\ll E_{cr}$, with  $E_i$ [$E_s$] being the electric field of IP [SP], then yield
\begin{equation}
\sqrt{\omega_i\omega_s} \xi_{i,s}\ll m\,.
 \label{eq:C3}
\end{equation}

The  electron and positron arise from vacuum in a certain phase of the SP, moving afterwards in the combined field of the IP and SP. The streak imaging is based on the signature of the initial phase of the SP in the electron (positron) energy exchange with the laser fields. The required preservation of this signature leads to the third condition, that the electron momentum is far from the resonance condition corresponding to the stimulated Compton process driven by the SP and IP: $\omega^{\prime}_i \gg \omega^{\prime}_s$, where $\omega^{\prime}_i=2\gamma_R\omega_i$, $\omega^{\prime}_s=\omega_s/2\gamma_R$ are the Doppler-shifted frequencies of the IP and SP in the electron rest frame, respectively, the Lorentz factor of the rest frame $\gamma_R$ is determined via $\omega^{\prime}_t=\omega_t/2\gamma_R=m_*$, and $m_{*}=m\sqrt{1+\xi_{i}^{2}/2}$ is the electron dressed mass. Thus, the off-resonance condition is
\begin{equation}
2\frac{\omega_i}{m}\frac{\omega_t^2}{m^2} \gg \frac{\omega_s}{m}\xi_i^2\,.
 \label{eq:C8}
\end{equation}

\section{The resolution}
\label{resolution}

To evaluate the resolution of SHEEP, we calculate via relativistic classical equations of motion the electron (positron) energy and momentum gain during the motion in the superposition of the IP and SP at the off-resonance condition.
The equations  for the transversal components of the electron momentum with respect to the laser propagation direction $z$ immediately follow from the canonical momentum conservation,
\begin{equation}
\label{pxpypz}
  p_x=q_x-eA_i(\eta)\,,\quad
  p_y=q_y-eA_s(\zeta)+eA_s(\zeta_0)\,,
\end{equation}
where $\eta=\omega_i(t-z)$ and $\zeta=\omega_s(t+z)$. The electron
is born at a phase $\zeta_0$ with drift momentum
$\textbf{q}=(q_x,q_y,q_z)$. From the Newton classical equations of
motion follows that the quantities $\Lambda\equiv {\cal E}-p_z$
and $\Pi\equiv {\cal E}+p_z$ obey the following equations
\begin{eqnarray}
    \frac{d\Lambda}{dt}&=&2e\frac{p_yE_s(\zeta)}{{\cal E}}\,, \\
    \frac{d\Pi}{dt}&=&2e\frac{p_xE_i(\eta)}{{\cal E}}\,,
\end{eqnarray}
where ${\cal E}$ is the electron energy. Due to the off-resonance
condition, there are two time scales in the electron dynamics,
fast and slow. Accordingly, when the independent variables
$\eta,\zeta$ are introduced,  a small parameter
$\epsilon=\omega^{\prime}_s/\omega^{\prime}_i=(\omega_s/\omega_i)(\Pi/\Lambda)\ll
1$ arises in the equations of motion 
\begin{eqnarray}\label{lambda_pi}
      \frac{\partial\Lambda}{\partial\eta}+\epsilon\frac{\partial\Lambda}{\partial\zeta}
      &=& -2\frac{\left[q_y+eA_s(\zeta_0)-eA_s(\zeta)\right]eA_s^{\prime}(\zeta)}{\Pi}\epsilon\,,\\
     \frac{\partial\Pi}{\partial\eta}+\epsilon\frac{\partial\Pi}{\partial\zeta}
      &=&-2\frac{[q_x-eA_i(\eta)]eA_i^{\prime}(\eta)}{\Lambda}\,.
\end{eqnarray}
We solve Eqs.~(\ref{lambda_pi}) by perturbation theory with respect to $\epsilon$. Additionally, the following initial conditions are used:  $\Lambda\rightarrow
\Lambda_0\equiv q_0-q_z$ upon switching off SP ($A_s \rightarrow 0$) and $\Pi
\rightarrow \Pi_0\equiv q_0+q_{z}$ upon switching off IP ($A_i \rightarrow 0$). After
the interaction with the IP and SP, the electron energy becomes
\begin{equation}
\label{energy}
    {\cal E}=q_0-\frac{m^2\xi_i^2}{4(q_0-q_z)}+\frac{q_yeA_s(\zeta_0)}{q_0+q_z}
    +\frac{e^2A_s^2(\zeta_0)}{2(q_0+q_z)}\,.
\end{equation}
The electron and positron are produced not only at the threshold
with zero momentum in the center-of-mass frame but also
above-threshold due to the possibility of surplus photon absorption
from the laser field. The number of absorbed IP photons at the threshold is $n_{i0}=m_*^2/\omega_t\omega_i$~\citep{footnote2}.
The width of variation of the absorbed laser photons ($n_i$) from the threshold value ($n_{i0}$)
 is of order $\delta n_i\sim n_{i0}$~\citep{Ritus}. Absorbing $n_i$ photons from the laser field, the particles in the center-of-mass frame are born with an energy  ${\cal E}_{cm}=\sqrt{n_i\omega_i\omega_t}$~\citep{footnote5}
and  with the polar emission angles  $\theta,\phi$ for the positron. The momenta  and energy of the particles in the lab frame  then are $p_{x0}^{\pm}=\pm\sqrt{\omega_tn_i\omega_i}\delta \sin\theta\cos\phi$, $p_{y0}^{\pm}=\pm\sqrt{\omega_tn_i\omega_i}\delta \sin\theta\sin\phi$ and ${\cal E}_{0}^{\pm}=(\omega_t+n_i \omega_i)(1\mp\beta_n\delta\cos\theta)/2$. Here, $\pm$ indices correspond to the  positron and electron, respectively, and $\delta\equiv \sqrt{\delta n_i/n_{i}}\lesssim 1/\sqrt{2}$, and $\beta_n\equiv  (\omega_t-n_i \omega_i)/(\omega_t+n_i \omega_i)\approx 1$. After the interaction with the laser fields ($A_i(\eta),A_s(\zeta) \rightarrow 0$), the momenta and energy of the particles are given by
\begin{eqnarray}
 p_{x}^{\pm}&=&\pm m_*\delta\sin\theta\cos\phi/\sqrt{1-\delta^2}\,, \\
p_{y}^{\pm}&=&\pm m_*\delta\sin\theta\sin\phi/\sqrt{1-\delta^2} \mp eA_s(\zeta_0)\,, \\
{\cal E}_{0}^{\pm}&\approx & \frac{\omega_t}{2}\left[ 1\mp\beta_n\delta \cos \theta +\frac{2\delta\sin\theta\sin\phi\sqrt{1-\delta^2}eA_s(\zeta_0)}{(1\pm\delta\cos\theta)m_*}
-\frac{m^2\xi_i^2}{2\omega_t^2(1\mp\delta\cos\theta)}+\frac{e^2A_s^2(\zeta_0)(1-\delta^2)}{(1\pm\delta\cos\theta)m_*^2}\right].
\end{eqnarray}
Note that 
the measurement of the positron energy in addition to that of the electron provides additional information whereas the positron transversal momenta do not, but the latter can be useful for a consistency check. Since the values $\{\theta,\phi,\delta,\zeta_0\}$ can be deduced from the measured $\{p_x,p_y,{\cal E}^+,{\cal E}^-\}$, the coincidence measurement of the electron and positron momenta after the interaction  provides  information on the pair production phase $\zeta_0$ in the SP.
\begingroup
\begin{table}
\begin{center}
\begin{tabular}{clccccc}
\hline\hline
& & \multicolumn{3}{c}{High energy TP} & \multicolumn{2}{c}{Low
energy TP} \, \tabularnewline & & Femto- & Atto- & Zeptosecond &
Atto- & Zeptosecond \tabularnewline \hline IP&
 $\begin{array}{c}
\\
\omega_{i} \:[\mathrm{eV}]\\
I_{i}\:[\mathrm{W}/\mathrm{cm}^2]\\
\xi_{i}\\
  {\mathcal N}_i
\end{array}$
&
$\begin{array}{c}
\\
 1 \,\\
10^{20}\\
10\\
 \sim 3
\end{array}$
&
$\begin{array}{c}
\\
 1 \,\\
10^{20}\\
10\\
\sim 3
\end{array}$
&
$\begin{array}{c}
\\
 1 \,\\
10^{20}\\
10\\
\sim 3
\end{array}$
&
 $\begin{array}{c}
\\
1000 \\
 10^{24}\\
1
\\ \sim 30
\end{array}$
&
 $\begin{array}{c}
\\
 1000 \\
 10^{24}\\
1
\\ \sim 30
\end{array}$
\tabularnewline
 \hline
SP&
$\begin{array}{c}
\\
\omega_{s} \:[\mathrm{eV}]\\
I_{s}\:[\mathrm{W}/\mathrm{cm}^2]
\\\xi_{s}\\
\end{array}$
& $\begin{array}{c}
\\
1\\
10^{18}
\\1
\end{array}$
& $\begin{array}{c}
\\
100\\
10^{22}
\\1
\end{array}$
& $\begin{array}{c}
\\
1000\\
10^{24}
\\1
\end{array}$
& $\begin{array}{c}
\\
100\\
10^{20}
\\0.1
\end{array}$
& $\begin{array}{c}
\\
1000\\
10^{22}
\\0.1
\end{array}$
\tabularnewline
\hline
TP&
$\begin{array}{c}
\\
\omega_{t}\:[\mathrm{GeV}] \\
\tau_{t} \:[as]\end{array}$
& $\begin{array}{c}
\\
 >30\\
10^2  - 10^3
\end{array}$
& $\begin{array}{c}
\\
 >30\\
1  - 10
\end{array}$
& $\begin{array}{c}
\\
 >30\\
0.1  - 1
\end{array}$
&
$\begin{array}{c}
\\
 >0.3\,\\
1-10\end{array}$
&
$\begin{array}{c}
\\
 >0.3\,\\
0.1-1\end{array}$
\tabularnewline
 \hline
\hline
\end{tabular}
\end{center}
\caption{SHEEP parameters for different combinations of intense
laser sources. $\Delta \omega_t/\omega_t\lesssim 0.1$,   and $N/S=10^{-2}$ are
assumed. $(N_{e+e-}/N_t)|_{\omega_t=\omega_{t \, min}}\sim 10^{-2}$ in all cases.  The XUV laser parameters can be realized in the ELI project~\citep{ELI}. \label{tab:comparison}}

\end{table}
\endgroup

The SHEEP resolution can then be estimated from the energy difference $\Delta\mathcal{E}$ of two electrons created at two different $\zeta_{1}$ and $\zeta_{2}$,
\begin{equation}
 \Delta\mathcal{E}\sim \omega_{t}\omega_{s}\tau_{t} \max\left\{\frac{\xi_s}{\sqrt{2}\xi_i},\frac{\xi_{s}^{2}}{\xi_i^2}\right\},
\label{deltaE}
\end{equation}
where the expressions $A_s(\zeta_{2})-A_s(\zeta_{1})\approx-E_s(\zeta_{0})(\zeta_{2}-\zeta_{1})/\omega_s$, $A_s^2(\zeta_{2})-A_s^2(\zeta_{1})\approx-2A_s (\zeta_{0})E_s(\zeta_{0})(\zeta_{2}-\zeta_{1})/\omega_s$ and $\zeta_{2}-\zeta_{1}=\omega_s\tau_t$ are used. 
The energy difference $\Delta\mathcal{E}$ due to streaking  
should exceed the energy uncertainty of the TP $\Delta\mathcal{E}\gg 1/\tau_{t}$ as well as the bandwidth $\Delta\omega_{t}$ of the $\gamma$-ray beam  $\Delta\mathcal{E}\gg \Delta\omega_{t}$. Using Eq.~(\ref{deltaE}) and assuming $\xi_i\gg \xi_s$, these conditions become
\begin{eqnarray}
\label{eq:streak}
(\omega_s\tau_t)^2 &\gg& (\omega_s/\omega_t)(\xi_i/\xi_s),\\
\Delta  \omega_t/\omega_t &\ll& \omega_s\tau_t(\xi_s/\xi_i).
\label{eq:streak2}
\end{eqnarray}

In a strong laser field, the electron dynamics will be disturbed by multiphoton Compton scattering. However, the probability of a photon emission in the multiphoton Compton process $W_C\sim \alpha \xi_i \mathcal{N}_i$ will be negligible when
\begin{equation}
\alpha \xi_i \mathcal{N}_i\ll 1,
 \label{C9}
\end{equation}
with the number of cycles in the IP $\mathcal{N}_i$. This
condition can be weakened to $\alpha \xi_i \mathcal{N}_i\sim 1$ by
selectively dropping Compton scattering events, which can be
identified by comparing momenta of the electron and positron after
the interaction. In the streaking regime  we have $\chi \sim 1$, and thus  with Eq. (10) $\alpha \xi_i \chi \ll 1$, while only in the opposite limit $\alpha \xi_i \chi \gtrsim 1$, the radiation dominated regime of multiphoton Compton scattering is entered~\citep{BulanovRDR,dipRR}. Similarly, a cascade of pair production \citep{Kirk, Narozhny} can only be initiated for $\chi \gtrsim 1$ if the interaction time $\tau_i=2\pi \mathcal{N}_i/\omega_i$ is  much larger than the pair creation time $\tau_{e^{+}e^{-}} \sim \omega_t/\alpha m^2 \chi^{2/3}$ \citep{Ritus}, which yields $\alpha \xi_i\mathcal{N}_i/\chi^{1/3}\gg 1$. But the opposite condition is fulfilled in the streaking regime and thus the pair production cascade is suppressed~\citep{footnote8}. 
Finally, basic preconditions for streak imaging are that the TP length
$\tau_{t}$ is shorter than half of the SP wavelength
$\lambda_{s}=2\pi /\omega_{s}$, and that the
streaking signal exceeds the noise level~\citep{Krausz2009},
\begin{equation}
\pi N/S \ll \omega_{s}\tau_{t}<\pi\,,
\label{eq:SN}
\end{equation}
where $S/N$ is the signal-to-noise ratio for the laser fields. The
resolution of the TP duration is directly related to the SP
frequency via this condition.  

\begin{figure}
\centering
\includegraphics[width=0.7\columnwidth]{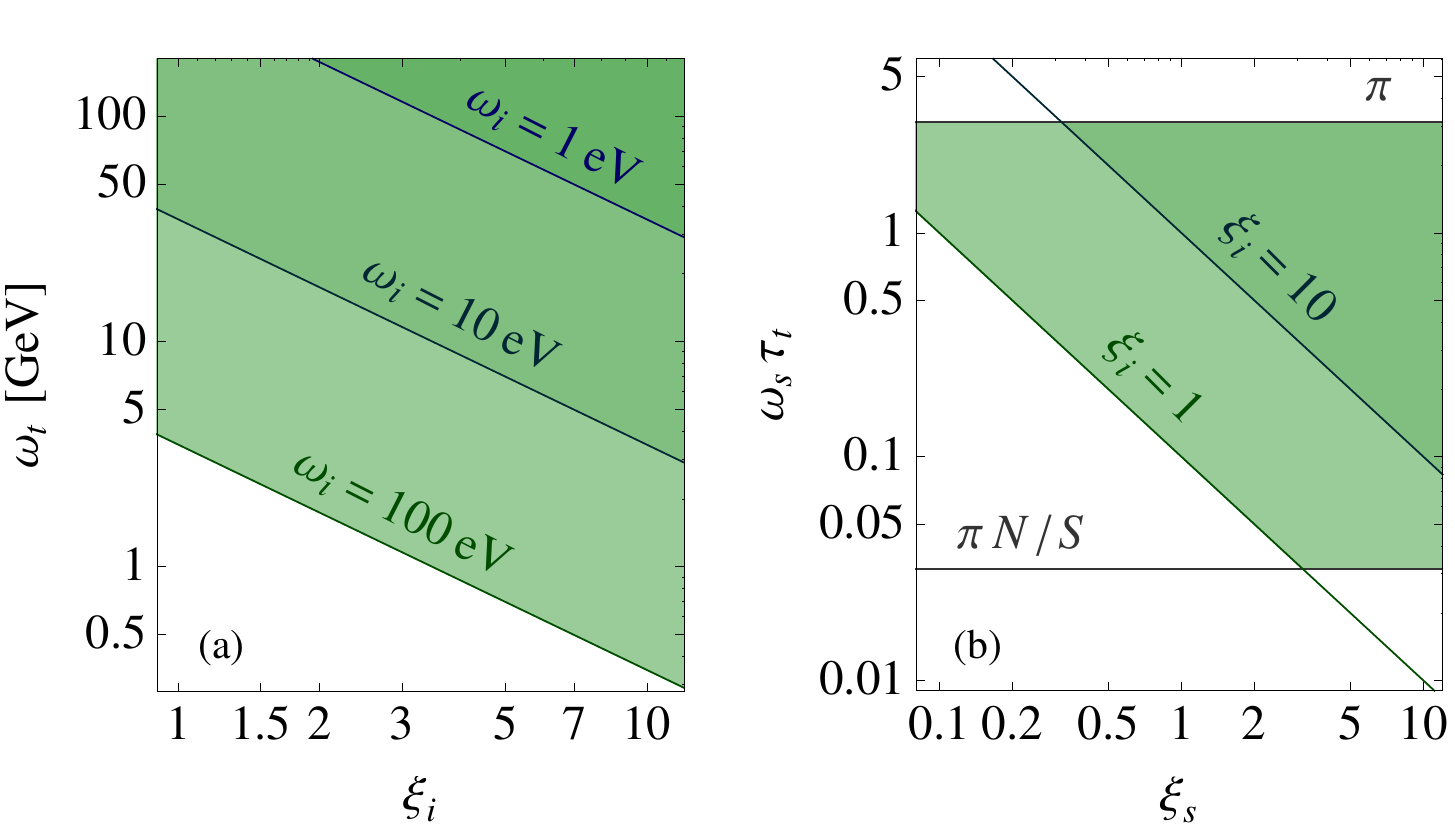}
\caption{(color online) Possible SHEEP ranges of (a) the TP
photon energy and (b) the TP duration. The allowed range of
$\omega_t$ in (a) (shaded with $\omega_i$-dependent hue) is mainly restricted by Eq.~(\ref{eq:C1})
and the range of $\tau_t$ in (b) (shaded with $\xi_i$-dependent hue) is mainly restricted by Eqs. (\ref{eq:streak2}) and (\ref{eq:SN}), for $\Delta \omega_t/\omega_t=0.1$ and $N/S=10^{-2}$.
}\label{fig:2}
\end{figure}

\section{The SHEEP parameters}
\label{parameters}

Table \ref{tab:comparison} shows a comparison of different
possibilities to realize SHEEP. The IP is a short and relatively
strong laser field with $\xi_i\sim 1-10$, ${\mathcal N}_i=3-30$ as
required from Eq.~(\ref{C9}). The minimal photon energy of the TP
depends on the IP frequency and intensity, given by Eq.~(\ref{eq:C1}), see  Fig.~\ref{fig:2}(a). Thus, at an infrared IP
with $\xi_i=10$ (corresponding to a laser intensity of
$I_i=10^{20}$ W/cm$^2$), one obtains $\omega_{t\,min}=30$ GeV,
while in the case of an ultraviolet IP with $\xi_i=1$
($\omega_i=1000$ eV, $I_i=10^{24}$ W/cm$^2$), instead
$\omega_{t\,min}=300$ MeV. We consider three regimes with SP of
different frequency: femtosecond TP with $\omega_s=1$ eV,
attosecond TP with $\omega_s=100$ eV and zeptosecond TP with
$\omega_s=1$ keV. The limitation on the minimal intensity of the
SP and on the TP resolution mainly arises from Eq.~(\ref{eq:streak2}) (see Fig.~\ref{fig:2}(b)), while the usual
streak condition Eq.~(\ref{eq:streak}) is easily fulfilled. If the
TP bandwidth is $\Delta \omega_t/\omega_t\sim 0.1$,
$\xi_s/\xi_i\gtrsim 0.1$ will be required. The required infrared
IP with an intensity of $10^{20}$ W/cm$^2$ is routinely available
in many labs. The intense high-frequency SP/IP with photon energies
in the $0.1-1$ keV range can be produced in the ELI facility via
high-order harmonic generation at plasma surfaces~\citep{ELI}. An
alternative realization could be provided by an XFEL if focusing
of x-rays becomes possible~\citep{footnote6}. 
The intensity of both TP and SP should be known with a precision determined by Eq.~(\ref{eq:SN})~\citep{footnote7}. Streaking requires detection of at least two electrons emitted from two different points in time within the TP. As  Table  \ref{tab:comparison} shows, this is possible  with hundreds of photons per TP.

\section{Conclusion}
\label{conclusion}

We have presented a detection scheme for the
characterization of short $\gamma$-ray pulses in the super-MeV 
energy range based on pair creation, facilitating a three beam setup of strong infrared and x-ray beams combined with the $\gamma$-ray test beam. Sub-attosecond time resolution could be achieved with high-order harmonic generation in the upcoming ELI facility.\\

We thank T.~Pfeifer for helpful discussions.

\bibliographystyle{elsarticle-num}
\bibliography{ipp_bibliography}

\begin{thebibliography}{10}
\bibitem{footnote2}
The energy and momentum conservation at the threshold of pair production in the
  center-of-mass frame of the produced particles yields
  $2\tilde{\gamma}_{cm}n_{i0}\omega_i=\omega_t/2\tilde{\gamma}_{cm}=m_*$, where
  $\tilde{\gamma}_{cm}$ is the Lorentz-factor of the center-of-mass frame and
  $n_{i0}=m_{*}^{2}/(\omega_{i}\omega_{t})$ the number of absorbed laser
  photons.
\bibitem{footnote8}
Note that with a SHEEP set up for the opposite condition as Eq.~(\ref{C9}),
  i.e.~$\alpha \xi_i \mathcal{N}_i\gg 1$, signatures of radiation reactions and
  cascade processes could also be tested.
\end{thebibliography}

\end{document}